# Low-temperature cotunneling electron transport in photo-switchable molecule-nanoparticle networks.

Yannick Viero[#], David Guérin, Dominique Vuillaume*.

*Institute for Electronics Microelectronics and Nanotechnology (IEMN), CNRS, Av. Poincaré, Villeneuve d'Ascq, France.*

*# Now at ESEO, 10 bd Jeanneteau, Angers, France.*

*\* E-mail: dominique.vuillaume@iemn.fr*

**ABSTRACT.**

We report the temperature-dependent (4.2 - 300 K) electron transport properties (current-voltage) of photo-switchable two-dimensional arrays of gold nanoparticles (10 nm in diameter) functionalized by azobenzene derivatives. Under UV-light irradiation at 4.2 K, the azobenzene moieties are switched from the trans to cis isomers, leading to an increase of the current. In both conformations, at low temperature (< 77 K) and low voltage (< 1 V) the voltage- and temperature-dependent current behaviors show that electron cotunneling is the dominant transport mechanism. The number of cotunneling events $N_{cot}$ slightly increases from ≈ 1.4 to 1.7 upon trans-to-cis isomerization of the azobenzenes. The nanoparticle Coulomb charging energy is not significantly modified (≈ 15 meV) by the azobenzene isomerization. This weak increase of $N_{cot}$ is explained by the modest cis/trans current ratio (≤ 10) and the limited numbers of nanoparticle-molecule-nanoparticle junctions inserted between the two nanoscale electrodes (< 50 nm apart) connecting the network.

# I. INTRODUCTION.

Two-dimensional arrays of metallic nanoparticles functionalized and/or linked by molecules (hereafter referred to as nanoparticles-molecules-networks (NMNs) are a powerful and versatile platform used in molecular electronics to study the basic electron transport and optical properties of molecules of interest and also assess the performances of several molecular electronics device applications, such as sensors, switches and plasmonic devices, and to assess their performances,[1-4] as well as to implement devices for unconventional computing like reconfigurable logic gates,[5-9] and neuro-inspired reservoir computing (RC).[9-12] The electron transport of NMNs at low temperatures was used to study the fundamental properties of mesoscopic charge transport. Depending on the nanoparticle size and density in the array, the nature and size of the molecular linkers, the degree of order/disorder in the networks, insulator-to-metal transition has been induced and different mechanisms of electron transport were studied in these NMNs.[13-20] In the case of weakly coupled nanoparticles and at low bias and low temperature, the inelastic cotunneling is the dominant mechanism.[13, 15, 19, 21] The cotunneling[22-24] is a mechanism whereby multiple, cooperative and synchronized tunneling events through the molecules between neighboring nanoparticles, are involved in optimizing the overall energy cost of electron transport in these systems. However, this cotunneling regime was only observed for a few types of molecules, mainly alkylthiols of various chain lengths,[13, 15, 17-19, 21, 25] and π-conjugated small molecules (oligo-phenylene-ethynylene-dithiol, biphenylpropanethiol).[21, 26] The role of the chemical structure was not clear (apart the evident increase of conductance for the π-conjugated molecules), the two systems (alkyl chains and oligo-phenylene-ethynylene) showing the same cotunneling fingerprints, i.e. the same number of cotunneling events $N_{cot}$ and temperature dependent behavior $N_{cot} \propto 1/T^{0.5}$.[21] Moreover, the effect of modifying the molecule conformation in situ inside the NMNs by applying an external stimulus was not reported. This last experiment is prone to shed more light on the relationship between cotunneling and molecular conformation.

Here, pursuing this objective, we report the study of cotunneling electron transport in photo-switchable NMNs made of gold NPs (10 nm in diameter) functionalized by azobenzene derivatives. Under UV-light irradiation, the azobenzene moieties are switched from the trans-to-cis isomers. At room temperature, we have previously shown that the current through the NMNs is increased upon the photo-isomerization to the cis form,[9, 27] and that the low-frequency noise (1/f$^n$ noise or flicker noise) is more pronounced with n increased from 1 to ≈1.4 (at low voltage).[12] In the present work, we focus on currents measured at low voltages (< 1 V) and low temperatures (4.2 - 77 K), where voltage- and temperature-dependent current behaviors show the fingerprint of cotunneling, namely power laws: $I \sim V^{2N_{cot}-1}$ and $I \sim T^{2N_{cot}-2}$ with $N_{cot}$ > 1. We observed a slight increase of $N_{cot}$ upon trans-to-cis isomerization from ≈1.4 to 1.8. The Coulomb charging energy of the nanoparticles is not significantly modified (≈ 15 meV) whatever the trans or cis isomers of the azobenzene capping the nanoparticles. These results extend the previously reported studies of NP networks functionalized by simpler molecules (alkyl chains and π-conjugated oligomers) and provide more insights toward the comprehension of the electron transport mechanisms in these molecular-based nanodevices. These findings will be hopefully valuable to optimize their properties and promote their use as a multifunctional platform in molecular electronics.

## II RESULTS AND DISCUSSION.

### A. Sample fabrication.

The azobenzenebithiophene (AzBT) molecules and the nanoparticle-molecules-networks (NMN) were fabricated as already reported in our previous works.[27, 28] In brief, gold nanoparticles (around 10 nm in diameter, Fig. S1 in the supplementary material)[27] were chemically capped with AzBT molecules (Fig. 1). We started with the synthesis of oleylamine capped Au NPs according to literature.[29] Then, we carried out ligand exchanges with the thiolated AzBT according to already reported recipes.[30] A monolayer of the AzBT capped NPs was deposited on e-beam Au lithographed electrodes (1 nm of titanium and 10 nm of gold) on a silicon wafer covered by 200 nm thick $SiO_2$. A Langmuir monolayer of the AzBT functionalized Au NPs was formed on a water meniscus surface[31] and transferred by dip coating on the substrate with the patterned electrodes. The presence and switching properties of the AzBT were checked by UV-vis spectroscopy (in solution) and XPS (on films) as detailed in our previous work.[27] The two needle-shaped face-to-face electrodes are separated by a gap of ≈ 30-50 nm (Fig. 1). The samples were mounted in a vacuum pumped (<$10^{-6}$ mbar) Lakeshore cryogenic multi-probe station equipped with a UV-vis transparent window for in situ light irradiation. The temperature dependent (4.2 to 300 K) current-voltage (I-V) curves were acquired with a Keysight 4156C semiconductor parameter analyzer. We first measured the I-V curves from 300 K down to 4.2 K for the trans-AzBT NMNs. About 50 samples were measured on the same chip at each temperature to assess the data distribution. Several I-Vs were discarded when no current is measured (no nanoparticle on the electrodes, since the coverage is not complete, Fig. 1) and when the two electrodes are short-circuited (high currents). Figure S2 (supplementary material) shows typical I-V distributions measured at 4.2 K for the trans-AzBT and cis-AzBT NMNs. The trans-to-cis isomerization was done at 4.2 K by irradiation with UV light (at 365 nm, xenon lamp) through the cryostat window for several hours. Then the temperature dependent I-V curves were acquired again from 4.2 K to 300 K for the cis-AzBT NMNs.

**B. Electrical properties.**

Figure 2 shows the temperature dependent current-voltage (I-V) curves recorded on NMNs with the AzBT in the trans and cis isomers (referred to trans-AzBT NMN and cis-AzBT NMN in the following). The cis-AzBT NMN shows a higher conductance (a factor ≈ 2 - 10) for the whole range of temperature (see Figs. S2 and S3, supplementary material). This cis/trans conductance ratio is consistent with our previously reported results for the same AzBT-NMN devices at 300 K.[9, 12, 27] The same datasets plotted in a log-log scale (Figs. 2c and d) reveal that the currents between ≈ 0.1 and 1 V follow a power law I ∝ $V^\alpha$, with an exponent α increasing when lowering the temperature. This behavior was systematically observed for all the measured samples (see Figs. S4 and S5 in the supplementary material) and it will be discussed below (*vide infra*, next section). To further analyze the temperature dependent behavior, these data (at 5 selected voltages) are plotted as ln(current) vs. 1 over temperature (Arrhenius plot) and current vs. temperature (in a log-log scale) in figure 3. For T ≤ 25 K, the current is not temperature dependent. For T ≥ 25 K, the Arrhenius plots show a thermally activated behavior with weak activation energies $E_A$ ≈ 6.2 to 9.3 meV (slightly voltage dependent as indicated in the figures) for the trans-AzBT NMN and ≈ 9.5 to 9.9 meV for the cis-AzBT NMN (Fig. 3a and 3c). Alternatively, Fig. 3b and 3d show that the currents also follow a power law, I ∝ $T^\beta$, with β ≈ 0.58 - 1.04 and 1.02 - 1.17 for the trans- and cis-AzBT NMNs, respectively. Both $E_A$ and β decrease when increasing the voltage. These behaviors are discussed in the next section.

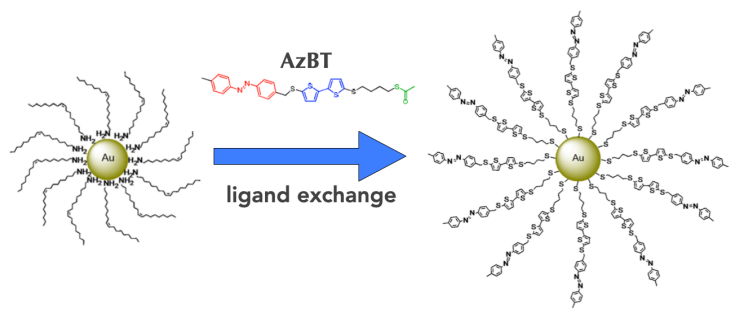

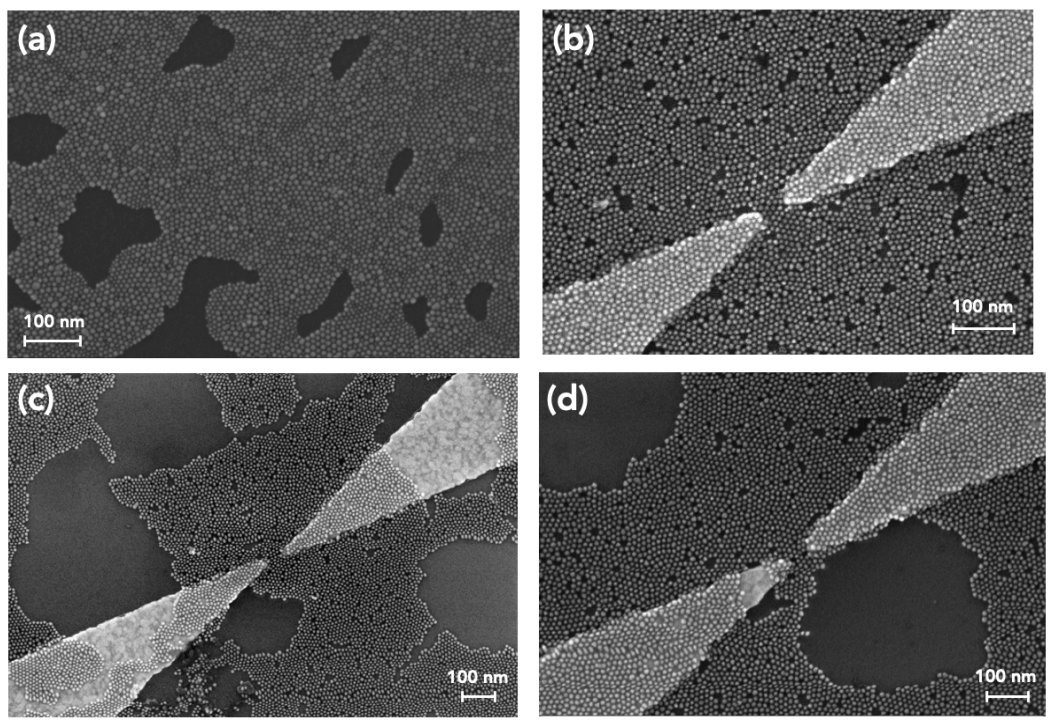

*Figure 1. Scheme of the synthesis of AzBT functionalized Au nanoparticles. Scanning electron microscope images: **(a)** NMN on a flat SiO$_2$ substrate (no electrode), **(b-d)** NMNs deposited on 3 different pairs of Au needle-shaped electrodes from the same chip. The electrode gap is ≈ 30-50 nm.*

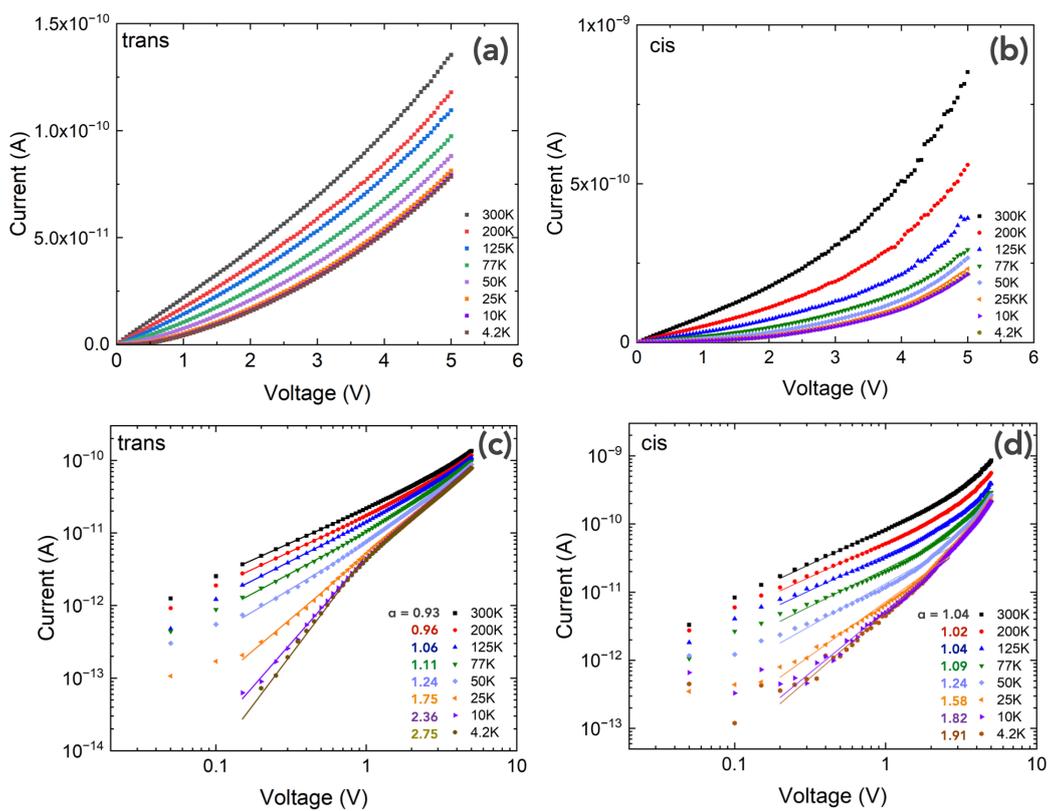

*Figure 2. Typical current-voltage (I-V) curves measured between 4.2 and 300 K: (a) for a trans-AzBT NMN and (b) for a cis-AzBT NMN. (c-d) Same data plotted in a log-log scale, showing a power law $I \propto V^\alpha$.*

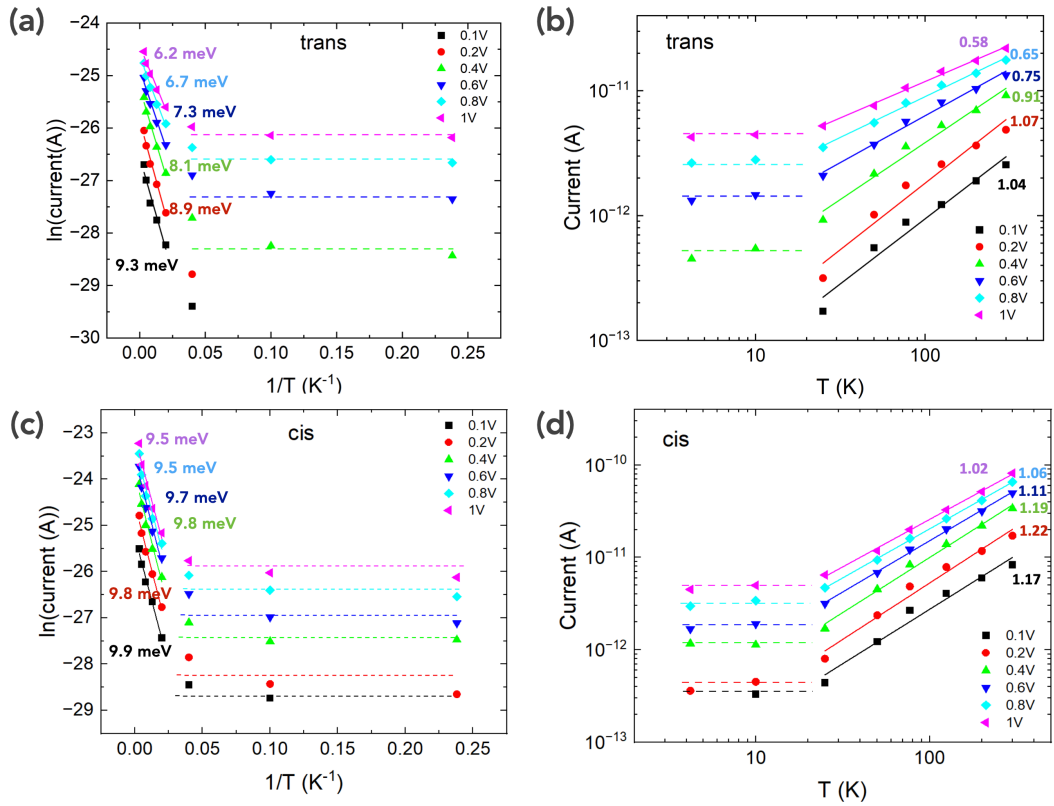

*Figure 3*. Temperature dependence of the currents (data from Fig. 2) at fixed voltages (low voltage 0.1 - 1V) for the trans-AzBT and cis-AzBT NMNs plotted as: *(a and c)* Arrhenius law ln(I) vs. 1/T and *(b and d)* power law I vs. $T^\beta$. For T ≥ 25 K, the activation energies $E_A$ and the exponent α are indicated in the graphs next to the plots with the same color code (solid lines are the fits). Below 25 K, the currents are almost constant, the dashed lines are guides for the eyes.

## C. Discussion.

The mechanisms leading to the increase of conductance in AzBT NMNs upon trans-to-cis isomerization have been previously described.[9, 27] For convenience, we briefly highlight the main reasons. As in self-assembled monolayers of AzBT contacted by a C-AFM tip, and supported by DFT calculations, the main reason of the conductance increase is due to the lowering of the energy barrier between the AzBT molecular orbital (here the lowest occupied molecular orbital) and the Fermi energy of the metal electrodes by about 0.4-0.5 eV.[32] Another factor is the organization of molecules in the gap between the NPs. In the NMNs, the electron transport between two adjacent NPs occurs through interdigitated AzBT molecules.[27] We have previously shown using molecular dynamics simulations that the number of interdigitated AzBT in the cis isomer are lower than in the trans one, thereby reducing the NMN conductance in the cis form and then the cis/trans conductance ratio.[27] This can explain the weak ratio (here 2-10, *vide supra*, and ≈ 30 in a previous study[27]) compared to the one measured on a self-assembled monolayer directly sandwiched between two electrodes (average value of $1.5 \times 10^3$, reaching up to $7 \times 10^3$).[32]

The voltage dependent power law of the current (I ∝ $V^\alpha$, Figs. 2, observed at low temperature, T < 77 K, and at low-voltage < 1 V) with α > 1 is the fingerprint of cotunneling electron transport in the NMNs.[13, 15, 21] In these temperatures and voltage conditions, the electron thermal energy is insufficient to overcome the Coulomb barrier (Coulomb blockade) of adding an electron to the NP, which is gauged by the charging energy $E_C$. Nevertheless, an alternative path is offered by the process of cotunneling. At larger T and V, the transport through the NMNs occurs by sequential, the electron energy is sufficient to overcome the Coulomb blockade and the electron transport in the NMNs occur by sequential tunneling through a series-parallel combination of NP-molecules-NP pathways. In this case, the current shows a weak temperature dependence (Fig. 2c and 2d at V ≳ 2 V). In the cotunneling mechanism, electron transfer occurs by cooperative electron tunneling through $N_{cot}$ single NP-molecule-NP molecular junctions in the NMN (inset Fig. 4). This process involves the creation of virtual intermediate states allowing electrons tunnel through the NPs without residing in the NP.[22] In this cotunneling regime, it was found experimentally and theoretically that the exponent α is given by α = $2N_{cot}$-1.[13, 15, 21] Figure 4 shows the $N_{cot}$ values calculated from all the I-V data (Fig. 2 and S4-S5). At 4.2 K, we have $N_{cot}$ ≈ 1.4 for the trans-AzBT NMNs and 1.4 - 1.7 for the cis-AzBT NMNs with a larger sample-to-sample dispersion (Fig. 4b). The temperature-dependent behavior of $N_{cot}$ partly follows the expected $1/T^{0.5}$ law[15, 21] (bold short dot lines in Figs. 4c and 4d, albeit with saturation observed at T < 10 K). Since the charging energy of AzBT capped Au NP is the key parameter separating the cotunneling and standard tunneling regimes in the NMNs, we estimated the charging energy $E_C$ from this temperature dependent behavior by[21]

$$N_{cot} = \sqrt{\frac{E_C}{kTln(e^2R_T/h)}} \qquad (1)$$

with k the Boltzmann constant, e the electron charge, h the Planck constant and $R_T$ the resistance of a single NP-molecule-NP junction. The values of $R_T$ are estimated from the I-V curves at 300 K (Fig. 2) and assuming that ≈ 3-5 NP-molecule-NP junctions are in series between the two electrodes (from SEM images). From the average slope of $N_{cot}$ vs. $1/T^{0.5}$ (Fig. 4, ≈2.1 and 2.2 for the trans and cis AzBT) and with $R_{T-trans}$ ≈ $5\times10^9$ - $2.5\times10^{10}$ Ω and $R_{T-cis}$ ≈ $5\times10^8$ -$5\times10^9$ Ω, we get $E_{C-trans}$ ≈ 4.6 - 5.2 meV and $E_{C-cis}$ ≈ 4.1 - 5.1 meV. These values appear weak compared to the expected and measured values of about 10 - 50 meV for NPs with a diameter of 10 nm.[17] A possible reason for this discrepancy is that Eq. 1 assumes a regular network with the same capacitance and resistance for every NP-molecule-NP individual junction,[21] while the real NMNs have a significant dispersion in size of the NPs and separation distances (see Fig. S1 and Ref. 27). Such a disorder induces deviations from the cotunneling model due to the distribution of the charging energy $E_C$ and the irregular shape and organization of the NPs.[33] Some deviations can also be induced by the difference between the thermal expansion of the NPs network and the underlying substrate, which can distort the NPs organization in the array.[18]

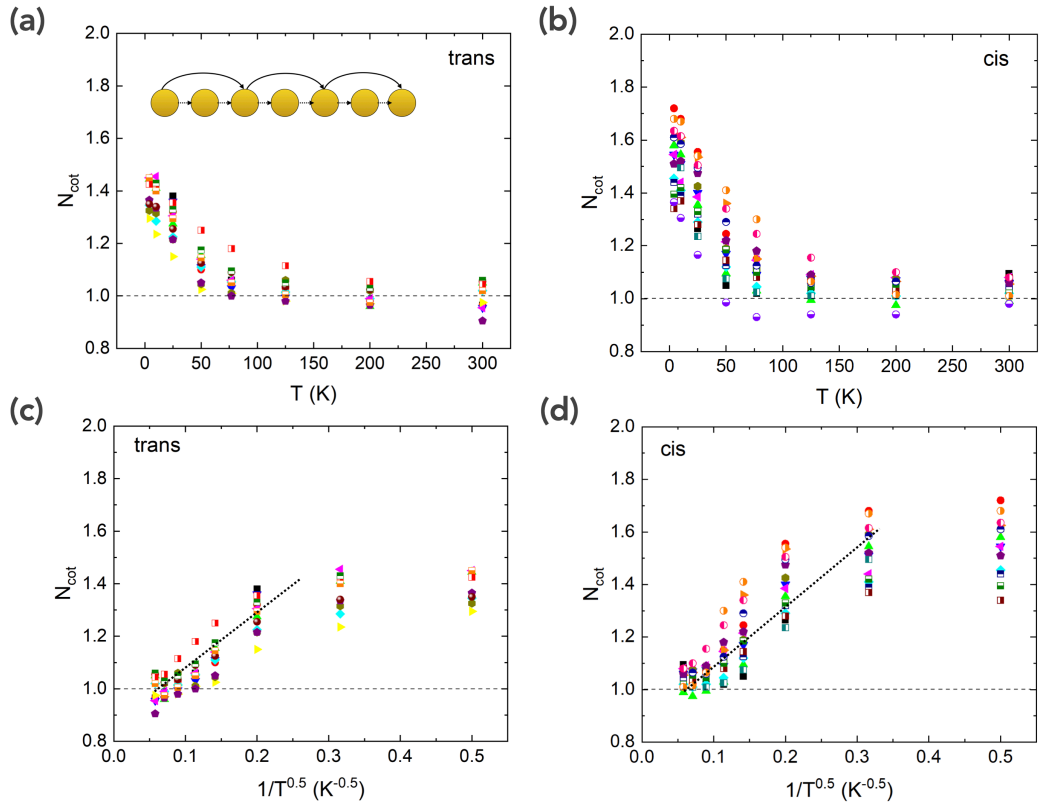

*Figure 4. (a-b)* Evolution of the cotunneling number $N_{cot}$ with the temperature for the trans- and cis-AzBT NMNs, respectively. *(c-d)* Same data plotted versus $T^{-0.5}$. The bold short dot lines show the linear dependence. The inset illustrates the cotunneling electron transport whereby cooperative tunneling jumps between adjacent NPs (small dotted arrows) allow electron transport on a larger distance between $N_{cot}$ junctions (here 2, solid bold arrows).

An estimation of the charging energy $E_C$ is also possible from a classical Arrhenius plot of the temperature-dependent currents (Figs. 3a and 3c). The activation energy, $E_A$, is plotted versus the voltage in Fig. 5a. According to the cotunneling model, the activation energy is related to the charging energy by:[15,21]

$$E_A = \frac{E_C}{N_{cot}} - eN_{cot}V_j \qquad (2)$$

where $V_j$ is the voltage across a single NP-molecule-NP junction in the NMN. The extrapolation at zero bias of the $E_A$ vs. V plots (Fig. 5a) give almost the same charging energy (assuming an average $N_{cot}$ of 1.5 in the two cases) $E_C \approx 14$ meV (trans-AzBT) and $\approx 15$ meV (cis-AzBT), a difference of the same order as the error bars. For Au NPs with a radius D (here $\approx 10$ nm) covered by a molecular layer of thickness t, the charging energy is given by[16, 17]

$$E_C = \frac{e^2}{4\pi\varepsilon_0\varepsilon_r}\left(\frac{1}{D} - \frac{1}{D+t}\right) \quad (3)$$

with $\varepsilon_0$ the vacuum permittivity and $\varepsilon_r$ the relative permittivity of the molecular layer capping the NP. Taken an average $t \approx 2.8$ nm (from ellipsometry measurements on self-assembled monolayers of the same AzBT molecules),[32] we deduced $\varepsilon_r \approx 3.5 - 3.8$, a reasonable value for many organic molecular monolayers as obtained by capacitance measurements,[34-37] and in agreement with a determination from the same approach on alkykthiol capped NP arrays.[13, 15] The slope of the $E_A$ vs. V curve (Fig. 5a) is significantly different for the trans (slope -3.51 meV/V) and the cis-NMNs (slope -0.46 meV/V). This slope depends on $N_{cot}$ and the fraction of the applied voltage seen by a single NP-molecule-NP junction, $V_j = \eta V$ (Eq. 2). Since $N_{cot}$ is also slightly depending on V (*vide infra*, Fig. 5b), we injected these $N_{cot}$ values at each V in Eq.2 and solved it to extract the value of $\eta$ (taken $E_C \approx 14$ meV (trans-AzBT) and $\approx 15$ meV (cis-AzBT), *vide supra*). We obtained $\eta = (4.40 \pm 0.17) \times 10^{-3}$ and $\eta = (5.8 \pm 1.8) \times 10^{-4}$ for the trans and cis conformation, respectively. This feature implies that a large fraction of the applied voltage is lost at the contact between the outer electrodes and the NP network. Otherwise, an ideal situation (no voltage drop at the contact) assuming a crude model $V_j = V/N_T$ with $N_T$ the total number of NP-molecule-NP single junctions between the two electrodes ($N_T \approx 5$, see Fig. 1) would have led to an electron energy in a single NP-molecule-NP junction $eV_j \gg E_C$ (in the investigated voltage range 0.1 - 1 V) and no possible observation of the cotunneling process in our NMNs. The variation of $\eta$ between the trans and cis conformations is difficult to explain without a detailed study on the structure at the interface between the NPs and the lithographed Au electrode (out of the scope of this work).

In the cotunneling regime, the temperature dependence of the current is also described by a power law for T > 25 K, $I \propto T^\beta$ (Figs. 3b and 3d), with $\beta = 2N_{cot}-2$.[15] The corresponding $N_{cot}$ values are shown in Fig. 5b. In agreement with the determination from the behavior of the I-V curves (Fig. 4), we observe a slight increase of the cotunneling for the cis-AzBT NMNs. Moreover, in the two cases, $N_{cot}$ slightly decreases when increasing the applied voltage, with a more important slope in the case of the trans-AzBT NMNs (Fig. 5b). To our knowledge, the previous works have not reported such a voltage-dependent behavior.[13, 15, 19, 21, 38] Increasing the applied voltage, the electron energy gradually increases, the cotunneling contribution tends to decrease and at large enough voltages, a sequential tunneling from NP to NP dominates.[15] If we extrapolate the curves in Fig. 5b, the value of $N_{cot}$ = 1 (sequential tunneling) is reached at V ≈ 2 V (trans conformation) and at V ≈ 5-6 V (cis conformation). We note that these values are in agreement with the voltages at which the I-V curves measured at different temperatures (Figs. 2c and 2d) tend to collapse to a rough temperature-independent behavior (sequential tunneling): around 2-3 V in Fig. 2c (trans-AzBT) and above 5 V in Fig. 2d (cis-AzBT).

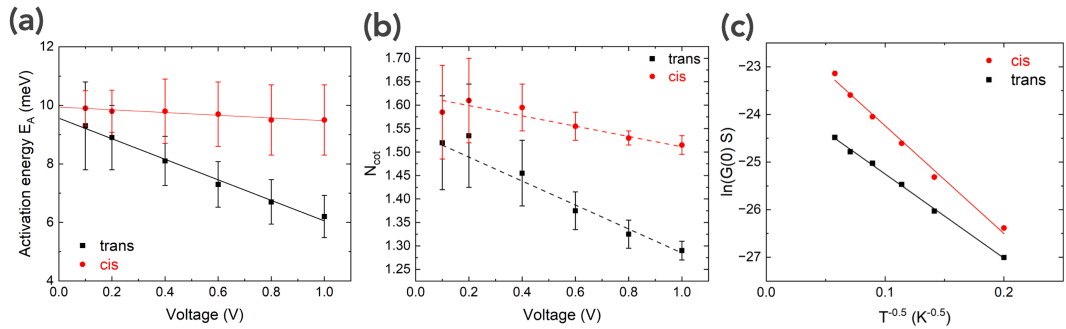

*Figure 5. (a) Voltage-dependent activation energy, $E_A$ determined by Arrhenius law (Figs. 3a and 3c). The solid lines are linear fits with a slope -3.51 meV/V (trans*

*isomer) and -0.46 meV/V (cis isomer).* ***(b)*** *$N_{cot}$ values calculated from the temperature-dependent power law of the currents (Figs. 3b and 3d). The dashed lines are linear fits with a slope -0.26 $V^{-1}$ and -0.11 $V^{-1}$ for the trans and cis isomers, respectively.* ***(c)*** *ln of the zero-bias conductance vs. $1/T^{0.5}$ (Eq. S4).*

The cotunneling mechanism also predicts that the zero-bias conductance (∂I/∂V calculated < 100 mV) follows an exponential inverse square-root temperature dependence[15]

$$G(0) \sim exp\left(-\sqrt{T_0/T}\right) \qquad (4)$$

similarly as the Efros-Shkolovskii variable range hopping in doped semiconductors[39] whereby the energy cost and tunneling events are balanced to optimize the global electron transport in the system. Figure 5c shows such a typical behavior for NMNs with AzBT in the trans or cis states. The fits (solid lines) with Eq. 4 gives a characteristic temperature $T_{0-trans} \approx 310$ K and $T_{0-cis} \approx 510$ K. These values are lower than those measured for alkylthiol and biphenylpropanethiol NMNs (> $10^3$ K).[13, 15, 19, 26] Since $T_0$ is given by $T_0 = CE_C/k\xi$ with $\xi$ the localization length of electrons and C a constant ≈ 2.8,[39] this implies a smaller $E_C$ or a larger $\xi$ for the AzBT NMNs. Since the NPs used in Refs. 13, 15, 19, 26 have a smaller diameter (4-5 nm), a smaller $E_C$ in our case is expected. In our case, assuming $\xi \approx 5$ nm ( $\xi < D$ in the case of weakly coupled NPs), we deduce from the $T_0$ values $E_C \approx 10$ and 15 meV, in good agreement with the determination from voltage and temperature dependent currents (*vide supra*).

Thus, these results show that the isomer conformation of the AzBT molecule (trans or cis) does not significantly modify the cotunneling process, which is characterized by the almost the same number of cotunneling event $N_{cot}$ and Coulomb charging energy for the two isomers. The values of $N_{cot}$ for the AzBT NMNs are smaller than the previously reported values up to 4-5,[13, 15, 17-19, 21, 25]

This feature might be a limitation due to the small number of available junctions (a maximum ≈ 5 in our NMNs compared to larger devices or devices hosting smaller nanoparticles) that impose a cutoff for the cotunneling mechanism.[15] The weak increase of $N_{cot}$ for the cis-AzBT NMNs may also be related to the modest increases of the overall conductance of the NMNs with respect to the trans-AzBT NMNs.

## III. CONCLUSION.

We have observed that electron cotunneling dominates the electronic transport properties (at low temperatures, below 77 K) of photo-switchable 2D arrays of molecularly (azobenzene derivatives) gold nanoparticles, confirming and extending previous findings for similar metal nanoparticle arrays functionalized with simpler molecules: alkylthiols and π-conjugated oligomers. In our nanoscale devices (connecting electrodes of the network with a gap length ≲ 50 nm), the number of cotunneling electron hoping is accordingly lower than in the previously reported larger devices (typically 100 nm and much more, up to the µm scale). We have also observed that the conformation, i.e. trans versus cis isomers, of the azobenzene shell has a negligible effect on the Coulomb charging energy of these molecularly functionalized gold nanoparticles.

## SUPPLEMENTARY MATERIAL.

See the supplementary material for details on statistical analysis of the NP sizes, statistical distribution of the current-voltage curves, additional data and complete data sets.

## ACKNOWLEDGEMENTS.

We acknowledge Sylvie Lepilliet for support with the low-temperature measurements, and Christophe Boyaval for help with the SEM measurements.

The study has been partly financially supported by ANR (grants ANR-12-BS03-2012 & ANR-12-BS10- 01801).

## AUTHOR DECLARATIONS.

**Conflict of interest.**

The authors have no conflicts to disclose.

**Author contributions.**

**Yannick Viero:** Data curation (equal); Formal analysis (equal); Investigation (lead); Ressources (equal); Validation( equal); Writing - review & editing (supporting).
**David Guerin:** Ressources (equal); Writing - review & editing (supporting).
**Dominique Vuillaume:** Conceptualization (lead); Data curation (equal); Formal analysis (equal); Funding acquisition (lead); Investigation (supporting); Supervision (lead); Validation (equal); Writing - original draft (lead); Writing - review & editing (lead).

## DATA AVAILABILITY.

The data that support the findings of this study are available from the corresponding author upon reasonable request.

## REFERENCES.

# Supplementary material

## Low-temperature cotunneling electron transport in photo-switchable molecule-nanoparticle networks.

Yannick Viero#, David Guérin, Dominique Vuillaume*.

Institute for Electronics Microelectronics and Nanotechnology (IEMN), CNRS,
Av. Poincaré, Villeneuve d'Ascq, France.

# Now at ESEO, 10 bd Jeanneteau, Angers, France.

* E-mail: dominique.vuillaume@iemn.fr

**Statistical analysis of the NP sizes.**

From the SEM images (Fig. 1 main text), we checked the size of the NPs (with ImageJ). Figure S1 shows the histogram of NP diameters, which can be fitted by two gaussian distributions. The main one gives an average diameter of 9.5 nm in good agreement with our previous work on the same NMNs.[1] The second one, with a mean diameter of about twice the NP diameter (19.7 nm) indicates that some nanoparticles are aggregated in the NMNs.

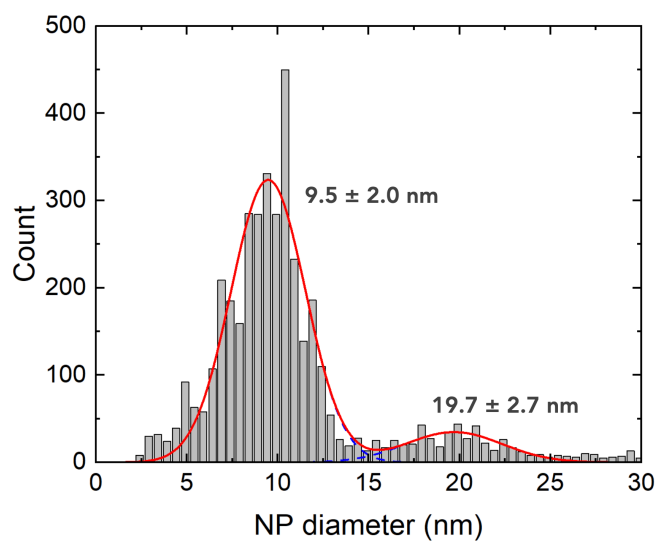

*Figure S1*. Histogram of the nanoparticle sizes fit by two gaussian distributions. The mean values are 9.5 and 19.7 nm and the standard deviations are 2.0 and 2.7 nm, respectively.

**Statistical distribution of the current-voltage curves and additional data.**

We plotted the I-V curves measured at 4.2 K for all the NMNs used during the course of this work with the AzBT in the trans and cis isomers. Albeit we observed a large dispersion in the both cases, the cis-AzBT NMNs display currents higher by a factor ≈ 5-10 (Figure S2). The same order of magnitude is observed for the 4.2 - 300 K temperature range (Figure S3).

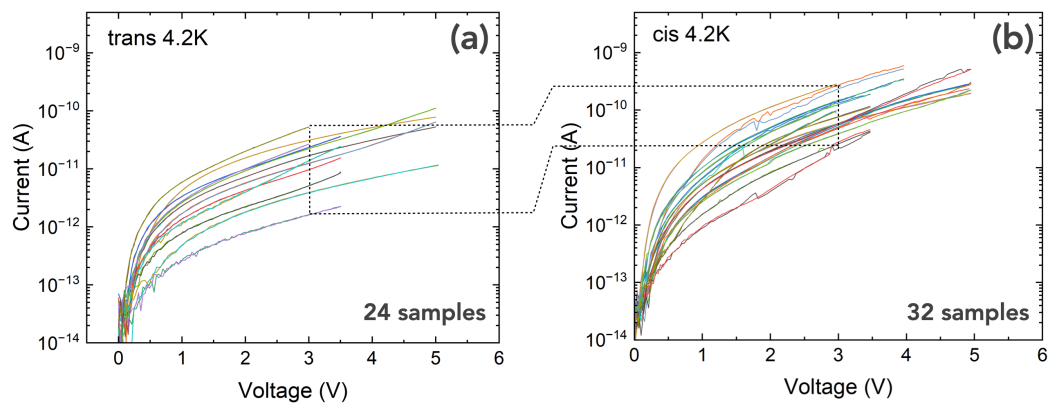

*Figure S2*. Current-voltage (I-V) dataset of several NMN samples measured at 4.2 K for the trans-AzBT and cis-AzBT NMNs.

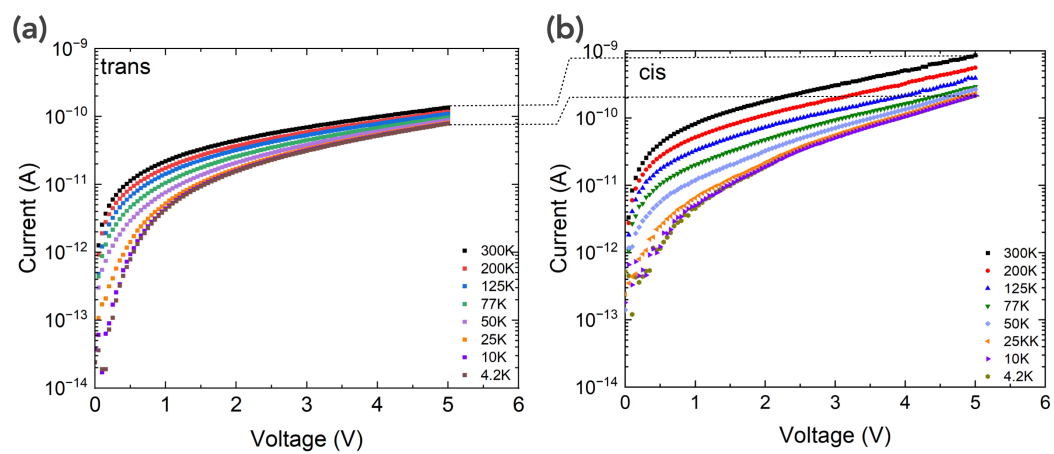

*Figure S3*. Typical I-Vs measured between 4.2 and 300 K with the AzBT in the two configurations (same data as figure 2, main text).

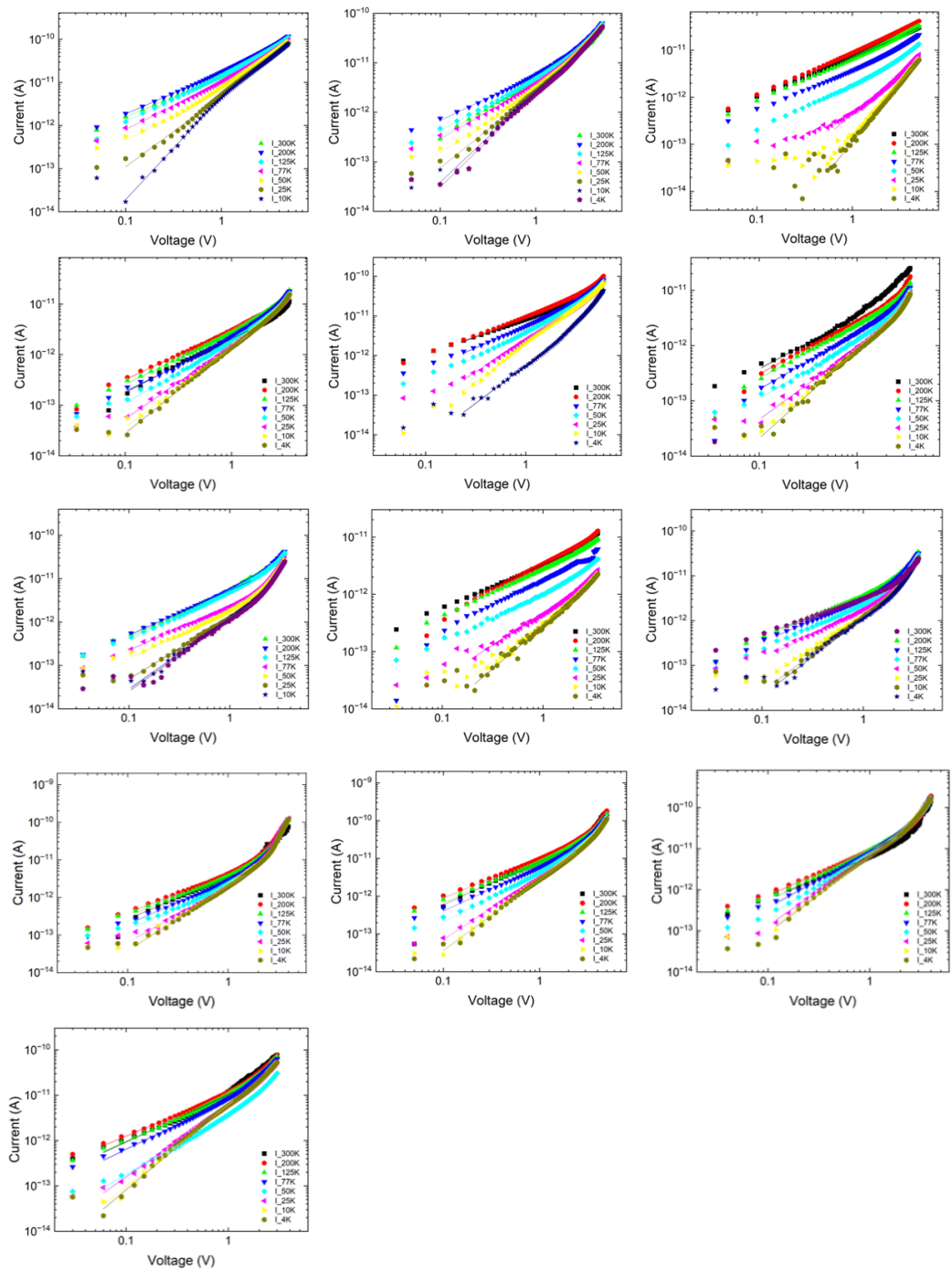

**F*igure S4*.** *All temperature-dependent I-V dataset (log-log scale) for the trans-AzBT NMNs.*

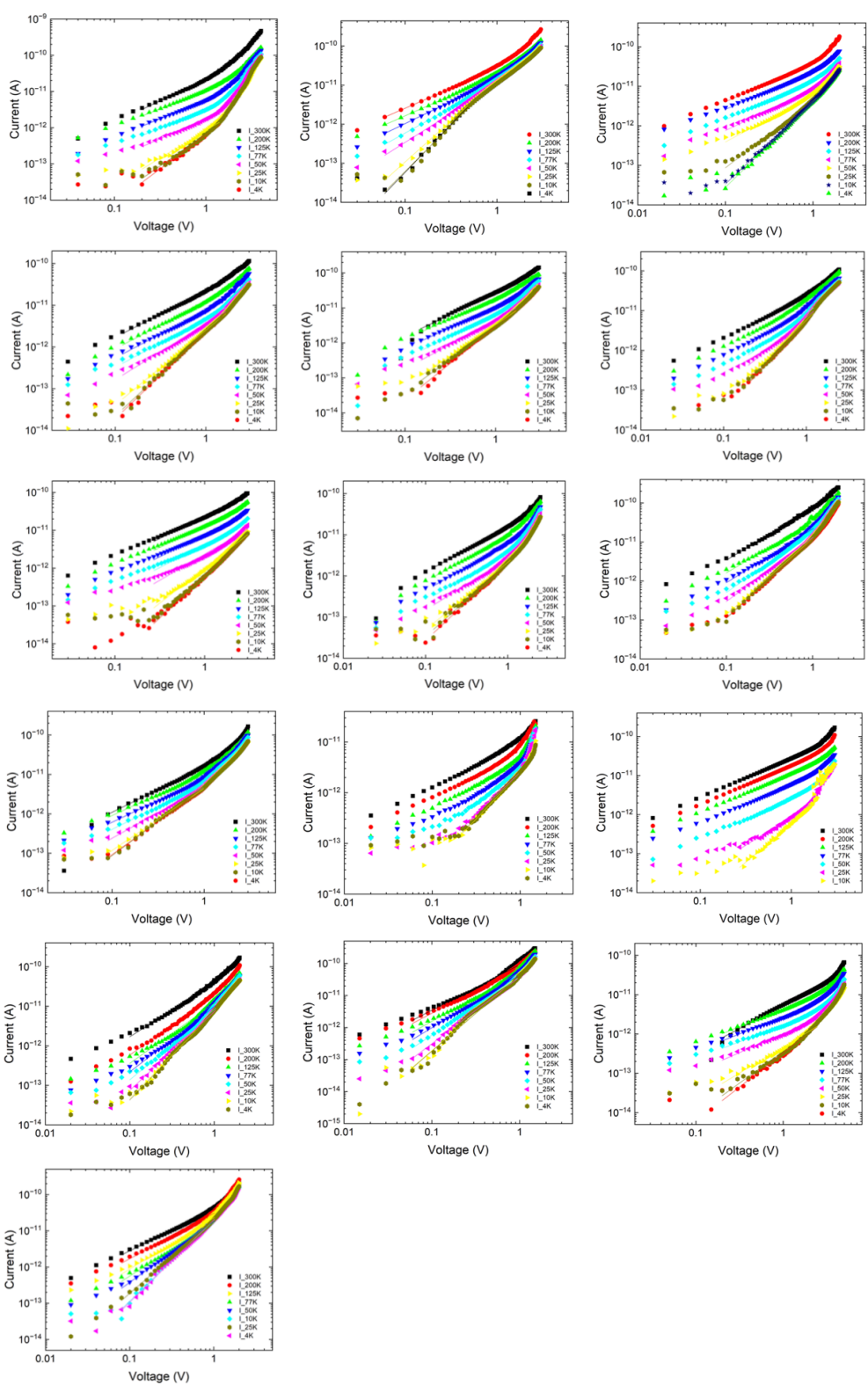

*Figure S5.* All temperature-dependent I-V dataset (log-log scale) for the cis-AzBT NMNs